\documentclass[aps,pra,twocolumn,superscriptaddress,showpacs]{revtex4}

\pdfoutput=1
\usepackage{graphicx}
\usepackage{amssymb}
\usepackage{amsmath}

\newcommand {\ket}[1] {|#1 \rangle}

\newcommand{\ueff}[0]{U_{\rm eff}^{(2)}}
\newcommand{\uefft}[0]{U_{\rm eff}^{(3)}}

\begin{document}

\title{Effective three-body interactions via photon-assisted tunneling in an optical lattice}

\author{Andrew J. Daley}
\affiliation{Department of Physics and Astronomy, University of Pittsburgh, Pittsburgh, Pennsylvania 15260, USA}
\affiliation{Department of Physics and SUPA, University of Strathclyde, Glasgow G4 0NG, UK}
\author{Jonathan Simon}
\affiliation{Department of Physics, University of Chicago, Chicago, Illinois 60637, USA}
\affiliation{Department of Physics, Harvard University, Cambridge, Massachusetts 02138, USA}

\date{November 6, 2013}


\begin{abstract}

We present a simple, experimentally realizable method to make coherent three-body interactions dominate the physics of an ultracold lattice gas. Our scheme employs either lattice modulation or laser-induced tunneling to reduce or turn off two-body interactions in a rotating frame, promoting three-body interactions arising from multi-orbital physics to leading-order processes. This approach provides a route to strongly-correlated phases of lattice gases that are beyond the reach of previously proposed dissipative three-body interactions. In particular, we study the mean-field phase diagram for spinless bosons with three- and two- body interactions, and provide a roadmap to dimer states of varying character in 1D. This new toolset should be immediately applicable in state-of-the-art cold atom experiments.

\end{abstract}

\maketitle

\section{Introduction}

Over the last decade, experiments with ultracold atomic gases in optical lattices have advanced to the point that many-body models can be engineered microscopically \cite{Bloch2008,Jaksch2005}. These developments have opened numerous opportunities to explore the properties of strongly interacting many-body systems under highly controlled conditions \cite{Lewenstein2007}. Part of the source of this fine control is that the dilute nature of the system leads to a strong dominance of two-body contact interactions over long-range or three-body processes in these systems, and especially three-body losses \cite{Jaksch1998}. However, there has recently been much interest in extending the breadth of quantum phases accessible with cold atoms by employing different types of interactions. In particular, three-body interactions should provide an opportunity to probe physics ranging from pair superfluidity of bosons \cite{Daley2009,Diehl2010a,Bonnes2011,Han2009,Lee2010,Diehl2010b,Diehl2010d} and fermions \cite{Kantian2009,Privitera2011,Titvinidze2011} to Pfaffian-like states \cite{Paredes2007,Roncaglia2010,Wojs2010} and modified Mott-insulator to superfluid phase-transition physics \cite{Silva-Valencia2011,Safavi-Naini2012,Sowinski2012,Luhmann2012}. Further ideas rely upon off-site interactions \cite{Capogrosso-Sansone2009,Chen2008,Zhou2010,ChenYC2011,Dalmonte2011,Ng2011}, which are potentialy realizable with polar molecules in optical lattices  \cite{Buchler2007b} or three-body constraints in effective bosonic models arising from spin-1 systems \cite{Mazza2010,Mahmud:2013aa}. Moreover, many proposed topological phases and spin liquids are constructed as ground states of Hamiltonians with three-body or other many-body interactions. Here we present a new method to engineer lattice models with dominant conservative three-body interactions for cold atoms in an optical lattice. 

Previous studies investigated achieving a three-body constraint via three-body loss, which suppresses triple occupation via a continuous quantum Zeno effect \cite{Daley2009}. The large on-site three-body loss rates suppress coherent tunneling processes that would otherwise form triply occupied sites, analogous to suppression of double occupation observed on a lossy Feshbach resonance \cite{Syassen2008,Garcia-Ripoll2009}. This effect is limited to producing a direct three-body constraint, i.e., it cannot be used to produce a finite three-body interaction, and comes at the cost of non-negligible three-body loss.  

At the same time, effective conservative three-body interactions in optical lattices have recently been observed in several different experiments \cite{Campbell2006,Will2010,Mark2011,Bakr2011,Mark2012}. These arise as shifts to normal on-site two-body interactions due to virtual population of higher Bloch bands, which are occupation-number dependent \cite{Johnson2009,Tiesinga2011,Dutta2011,Bissbort2011}. Because these arise from virtual couplings to higher bands, the energy shift $\delta U_3$ is naturally much smaller than the onsite interaction shift for two particles $U$. Here we discuss an approach that amplifies the role of $\delta U_3$ by using photon-assisted tunnelling (e.g., a Raman process or lattice modulations) to coherently provide an energy $\omega_m \approx U$, necessary for an atom to tunnel onto a site already occupied by another atom. If we begin with a large $U$, then the resulting effective two-body interactions will be the detuning of the laser-assisted tunneling process, $\ueff \approx \omega_m-U$, and these two-body interactions can be made much smaller than $\delta U_3$.  If we begin in a limit where $U$ is very large, then $\delta U_3$ can be much larger than tunneling rates in the lattice, and three-body interactions can be the dominant energy scale of the effective model. This approach is similar to Floquet hamiltonian schemes proposed to generate synthetic gauge fields for both cold atoms \cite{Kitagawa2010} and solid-state systems \cite{Lindner2011}.

Below we discuss this mechanism in detail, and illustrate the new opportunities by considering a system of bosons in an optical lattice with a finite three-body interaction.  We discuss the corresponding phase diagrams, as well as observability and preparation of different phases of dimers in these systems \cite{Daley2009,Diehl2010a,Bonnes2011}. We also discuss the preparation of metastable many-body phases via adiabatic processes, which is enhanced because our scheme provides a means to directly control the effective two-body interactions $\ueff$ by adjusting the frequency $\omega_m$ relative to $U$. This degree of control for adiabatic state preparation is particularly important, because in the most straightforward experimental implementations, our scheme will lead to a large attractive three-body interaction. However, under the right circumstances, large attractive three-body interactions can suppress triply occupied sites to the same extent as repulsive three-body interactions. As an example, we consider the dynamics of time-dependent ramps that could be used to realize many-body states of dimers in 1D. 

The remainder of this paper is organized as follows: In section \ref{sec:engineering} we describe the procedure to enhance the role of three-body interactions using laser assisted tunneling or lattice modulation. In sections 
\ref{sec:mfpd} -- \ref{sec:prepare} we then give examples of many-body physics that is made accessible by this scheme, beginning with section \ref{sec:mfpd}, where we derive a mean-field phase diagram for the Bose-Hubbard model with finite three-body interactions. In section \ref{sec:dimerint} we discuss controlling dimer-dimer interactions for attractive two-body interactions via the effective three-body interaction, and in section \ref{sec:prepare}, we consider time-dependent preparation of many-body states, which are typically metastable in this system. This is illustrated by analyzing preparation of the states discussed in section \ref{sec:dimerint}. In section \ref{sec:summary} we provide a summary and outlook. 

\section{Engineering $3$-body interactions via modulation or laser assisted tunneling}
\label{sec:engineering}

The three-body interactions we consider here arise physically because of shifts in the bound state energies corresponding to different numbers of particles on a lattice site \cite{Campbell2006,Will2010,Mark2011,Bakr2011,Mark2012,Johnson2009,Tiesinga2011,Dutta2011,Bissbort2011}. In their simplest form these corrections arise due to interactions coupling the system virtually to higher Bloch bands \cite{Johnson2009}. This can be seen intuitively, e.g., as a broadening of the on-site wavefunction for repulsive interactions, which reduces the effective interaction energy for larger $n$. The standard Bose-Hubbard Hamiltonian describing bosonic atoms in the lowest band of a deep optical lattice can be modified to reproduce the number-dependent bound state energies \cite{Johnson2009,Will2010,Mark2011} as ($\hbar\equiv 1$)
\begin{eqnarray}
H_{\rm MBH}&=&-J \sum_{\langle i,j\rangle}b^\dagger_i b_j + \frac{U}{2}\sum_i \hat n_i (\hat n_i -1) \nonumber \\
& &+ \frac {\delta U_3}{6} \sum_i  \hat n_i(\hat n_i-1)(\hat n_i-2) \nonumber\\
& &+\sum_{N>3} \delta U_N \sum_i  (b^\dagger _i)^N (b_i)^N +\sum_i \varepsilon_i \hat n_i,\label{originalham}
\end{eqnarray} 
where $b_i$ is a standard bosonic annihilation operator for an atom on site $i$, $\hat n_i = b^\dagger_i b_i$, $\langle \ldots \rangle$ denotes a sum over neighboring sites, $J$ is the tunneling amplitude between nearest neighbor sites, $\varepsilon_i$ is the energy shift at site $i$ due to external trapping potentials, $U$ is the onsite interaction energy shift for two particles on a site, and $\delta U_3$ is the correction to this energy shift when three particles are present on a lattice site. Similar terms for $N>3$ particles are included with shifts proportional to $\delta U_N$. For atoms in a Harmonic trap, $\delta U_3$ can be estimated as $U/(1+U/\omega_g)-U$, where $\omega_g$ is the oscillator spacing \cite{Johnson2009, Mark2011, Mark2012}. We note that while shifts in this form typically give rise to $\delta U_3<0$, the alternative of introducing an RF coupling to a three-body bound state \cite{Safavi-Naini2012} could give rise to 3-body shifts of either sign.

Some care must be taken in using this model, as the single band Bose-Hubbard model is only valid in the limit $U \bar n /\omega_g \ll 1$, with $\bar n$ the mean density, where $\delta U_N$ only makes very small modifications to the system dynamics. As shown in Ref. \cite{Buchler2010}, once off-resonant coupling to higher bands is important and thus $\delta U_N$ significant, the single band Bose-Hubbard model will not be able to simultaneously reproduce the scattering amplitude and bound state energies for two particles in an optical lattice. Here we will use this model only in the limit where $U\gg J$, and consider coupling between neighboring lattice sites that relies on an accurate description of the bound state energies. We will also assume that all parameters other than $U$ are much smaller than $\omega_g$, so that higher bands are only virtually populated and their effect incorporated through interaction shifts $\delta U_N$ as written above.

\begin{figure}[tb]
\includegraphics[width=8.5cm]{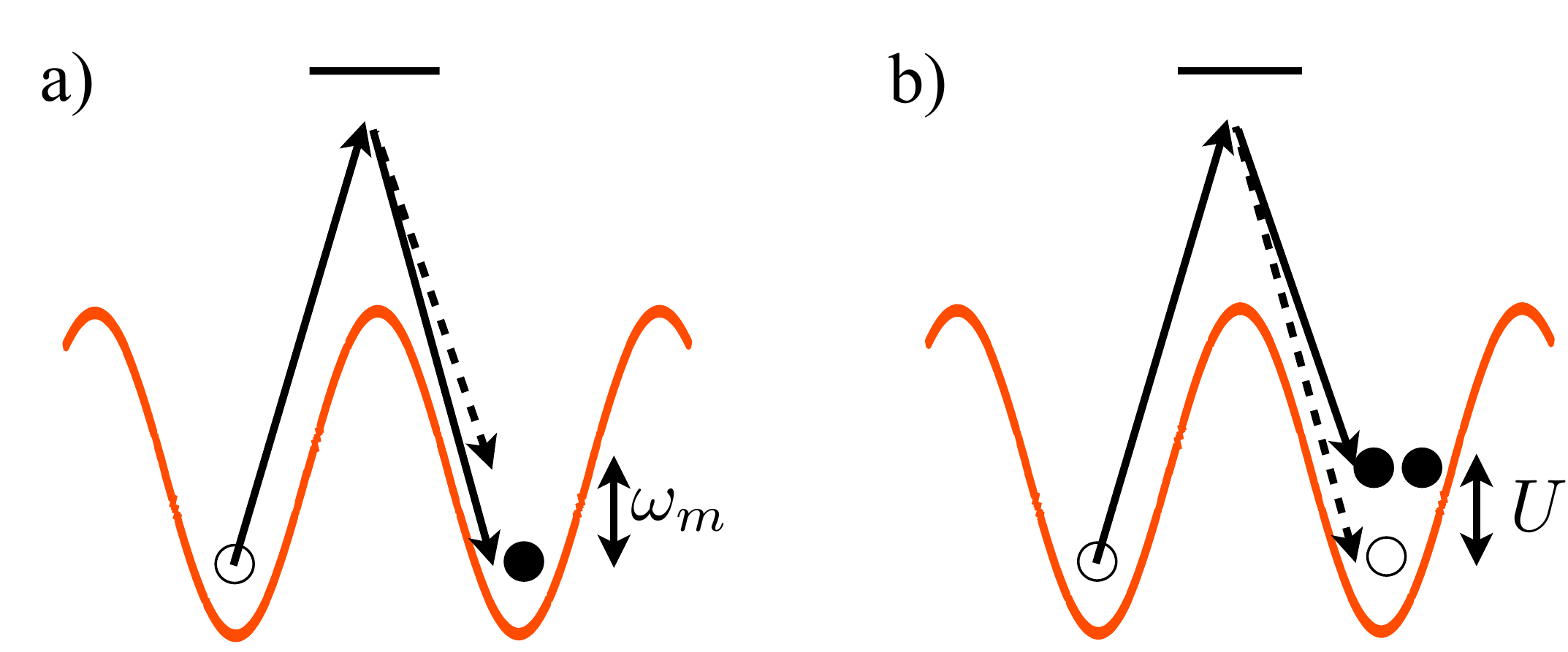}
\caption{Schematic of the laser-assisted tunneling scheme that converts three-body shifts into dominant three-body interactions, in a strongly interacting system, $U\gg J$. (a) By adding a Raman process with frequency $\omega_m\sim U$, we do not affect tunneling where an atom in a singly occupied site tunnels into an unoccupied site (or the equivalent process where an atom from a doubly occupied site tunnels onto a site with a single particle present). However, (b) the sideband gives rise to near-resonant couplings when an atom in a singly occupied site tunnels so as to produce a doubly occupied site. For $U\gg J$ This results in an effective model where $U_{\rm eff}^{(2)} = \omega_m-U$.} \label{fig:laserassisted}
\end{figure}

Below we outline two schemes to make $\delta U_3$ the dominant interaction scale in the system dynamics. The basic principle for both schemes is sketched in Fig.~\ref{fig:laserassisted}. We work in the limit described above, where $U\gg J$, and engineer a time-dependent $J$ with a modulation frequency $\omega_m \sim U$, of the form $J(t)=J_0 + J_1 \cos(\omega_m t)$. In the case of a laser assisted tunneling produced by a Raman coupling, this can be produced by choosing a Raman detuning $\omega_m$, and in the case of a modulated lattice, this time-dependence of the tunneling can be directly engineered by varying the lattice depth. In each case, when we consider the resulting effective tunneling process, we see that where a particle in a singly occupied site tunnels resonantly onto an empty neighbouring site with amplitude $J_0$ and the modulated part of the coupling plays little role, as it is far detuned from the transition when $\omega_m \sim U \gg J_0,J_1$. This is depicted in Fig.~\ref{fig:laserassisted}a. When a particle in a singly occupied site tunnels onto an initially singly-occupied neighbouring site, the coupling will have amplitude $J_1$, as the unmodulated coupling is far from resonance, as shown in Fig.~\ref{fig:laserassisted}b. This creates an occupation-dependent tunneling amplitude, and for a lattice system with singly and doubly occupied sites only, gives rise to an effective two-body interaction determined by the detuning of the modulation, $\ueff=-\Delta=\omega_m-U$. This can be extended to work for occupations up to three particles on any lattice site by the introduction of a second modulation frequency, $J(t)=J_0 + J_1 [\cos(\omega_{m1} t)+\cos(\omega_{m2} t)] $, with $\omega_{m2}=2U+\Delta$, that provides the coupling $J_1$ with the same detuning and effective two-body interaction $\ueff$ when an atom in a singly occupied site tunnels onto a site that is already doubly occupied. This effective two-body interaction, $\ueff$ can now be made positive or negative by tuning $\Delta$, giving us a new way to flexibly control the two-body interactions. Moreover, this interaction can be made much smaller than $\delta U_3$, which results in dominant, real three-body interactions in the system. We note that in order to ensure that the approximation that we essentially have at most three atoms per site remains valid, we require that $\delta U_3 \gg \ueff$ if we enter a regime where $\ueff=-\Delta \sim J_1$.

Below we analyze this scheme in detail for implementations with laser-assisted tunneling and lattice modulations. We first begin by considering a two-site system, and then write down an general effective lattice model, which is valid in the limit $J\ll U \sim \omega_m \ll \omega_g$.

\subsection{Laser-assisted tunneling}

The first realization of this scheme can be achieved with laser-assisted tunneling, which has recently been used to implement gauge fields in experiments \cite{Aidelsburger:2013aa,Miyake:2013aa}. A related scheme was also recently suggested as a means to produce an occupation-dependent tunneling phase, leading to anyonic statistics in 1D \cite{Keilmann2011}. The idea is to produce a Raman process at two different frequencies, corresponding to energy differences of $\omega_{m1}$ and $\omega_{m2}$ \footnote{Note that Raman scheme can also modify effective interactions via number-dependent AC stark shifts. These can be incorporated in a redefinition of the original parameters in Eq.~\ref{originalham}.}, which can provide the necessary energy to particle as they tunnel. Note that our scheme does not require change in an internal spin state, and so can be detuned far from resonance, thus reducing heating from light scattering.

In combination with standard tunnelling (which will remain the dominant process for a particle tunnelling resonantly between two sites), the Raman processes give rise to the effective tunneling 
\begin{equation}
J(t)=J_{0}+J_1\left(e^{i\omega_{m1}t}+e^{-i\omega_{m1}t}\right)+J_1\left(e^{i\omega_{m2}t}+e^{-i\omega_{m2}t}\right),
\end{equation}
with $\omega_{m1}=U+\Delta$ and $\omega_{m2}=2U+\Delta$ as defined above. If we assume that the lattice is separable and consider tunnelling along one lattice axis, $\mathbf{e}_x$, then for small coupling, we note that
\[J_1\approx \int \,dx\, w(x) {\rm e}^{i\mathbf{\delta k . e}_x x} w(x-a),\] with $\mathbf{\delta k}$ the difference in wave numbers for the two arms of the Raman transition, $w(x)$ the Wannier function in the lowest Bloch band along direction $\mathbf{e}_x$, and $a$ the corresdponing lattice spacing.

We will now consider the effects of this time-dependent tunneling on a two-site system. We will neglect corrections for higher occupation numbers than 3 atoms per site, which is valid provided $U\gg J$ and that the filling factor
on the lattice is sufficiently small. In the limit $\omega_g \gg U\gg \delta U_{3},J_{0},J_1$, the Hilbert space of the system with an unspecified total particle number is divided into different manifolds
with different expectation values $E_{U}=\langle H_{U}\rangle$ of
the hamiltonian $H_{U}=(U/2)\sum_{i}\hat n_{i}(\hat n_{i}-1)$. As $U$ the largest energy scale in the problem once we restrict
to the lowest Bloch band, these manifolds are coupled only by the oscillating terms proportional to $J_1$, as other processes
are far off resonance and can be removed in perturbation theory. Within each manifold, the degeneracy of states is partially lifted by tunneling terms proportional to $J_{0}$ that couple states inside each manifold. We can then characterize the tunneling in a basis of states with fixed occupation numbers $(n_{L},n_{R})$ on each site, by calculating the coupling strength and detuning from the nearest coupling resonance between each combination of states. Computing the final and initial energy expectation values of $H_U$,$E_{U}^{(f)}$ and $E_{U}^{(i)}$, these can be tabulated for up to four-particle occupations as:
\begin{center}
\begin{tabular}{|c|c|c|c|c|}
\hline 
Initial & Final & $E_{U}^{(f)}-E_{U}^{(i)}$ & Detuning  & Coupling\tabularnewline
\hline 
\hline 
(0,1) & (1,0) & 0 & 0 & $J_{0}$\tabularnewline
\hline 
(1,1) & (2,0) & $U$ & $\Delta$ & $J_1$\tabularnewline
\hline 
(1,2) & (2,1) & 0 & 0 & $J_{0}$\tabularnewline
\hline 
(1,2) & (0,3) & $2U$-$\delta U_{3}$ &  $\Delta-\delta U_{3}$ & $J_1$\tabularnewline
\hline 
(2,2) & (3,1) & $U$$-\delta U_{3}$ & $\Delta-\delta U_{3}$ & $J_1$\tabularnewline
\hline 
(1,3) & (0,4) & $3U+\delta U_{4}-\delta U_{3}$ & - & -\tabularnewline
\hline 
(2,3) & (3,2) & 0  & 0 & $J_{0}$\tabularnewline
\hline 
(2,3) & (1,4) & $2U+\delta U_{4}-\delta U_{3}$ & $\Delta-U_{4}+\delta U_{3}$ & $J_1$\tabularnewline
\hline 
(3,3) & (4,2) & $U+\delta U_{4}-2\delta U_{3}$ & $\Delta-U_{4}+2\delta U_{3}$ & $J_1$\tabularnewline
\hline 
\end{tabular}
\par\end{center}

We note immediately that the energy differences can now be characterized in terms of the effective two-body interaction $\ueff=-\Delta$, which is controllable via
the modulation frequencies $\omega_{w1}$ and $\omega_{w2}$, and the three-body interaction $\uefft=\delta U_{3}$. In a typical experiment, we would expect that $|\delta U_4|>|\delta U_3|$, and hence neglecting four particle occupation will typically be a good approximation in the limit where $\delta U_3 \gg J_0,J_1$. In the limit $J_{0},J_1,|\delta U_{3}|,|\Delta|)\ll U$ we can now make a rotating wave approximation, neglecting coupling terms in the Hamiltonian that rotate with frequencies of the order of $U$. This gives us the effective model
\begin{eqnarray}
H_{\mathrm{eff}}&=&\frac{U_{\mathrm{eff}}^{(2)}}{2}\sum_{i}n_{i}(n_{i}-1)+\frac{U_{\mathrm{eff}}^{(3)}}{6}\sum_{i}n_{i}(n_{i}-1)(n_{i}-2)\nonumber \\
& &-J_{0}'\sum_{\langle i,j\rangle,n}b_{i}^{\dagger}b_{j}P_{i=n,j=n+1}-J_1\sum_{\langle i,j\rangle}b_{i}^{\dagger}b_{j}, \label{effectiveham}
\end{eqnarray}
where $P_{i=n_{1},j=n_{2}}$ is projector on states with particular
on-site particle numbers for the sites $i$ and $j$, and $J_{0}'=J_0-J_1$.

In the limit $J_0,J_1 \ll U$, and with a homogeneous initial system $\varepsilon_i=0$, this Hamiltonian also constitutes an effective model for a multiple-site system. We note that position-dependent lattice depths, caused by the finite beam waist of the laser light creating the lattice will give rise to a shift in both the on-site potential $\varepsilon_i$ and the interactions
$U$. These will lead to a spatially-varying shift in the effective detunings, and thus the final values of $U_{\mathrm{eff}}^{(2)}$ and $U_{\mathrm{eff}}^{(3)}$, which will become important when the variation from site to site of $\varepsilon_i$ becomes comparable to  $U_{\mathrm{eff}}^{(2)}$ and $U_{\mathrm{eff}}^{(3)}$. This should play only a small role when $U_{\mathrm{eff}}^{(2)}, U_{\mathrm{eff}}^{(3)}\gg J_0,J_1$, but in the limit where the tunneling dominates, this technique provides a potentially interesting means to engineer spatially-dependent interactions. 

\subsection{Lattice modulations}

An equivalent method to generate a time-dependent tunneling $J(t)$ is to modulate the depth of the optical lattice potential. This is mathematically equivalent to Raman-induced processes, with a choice of the wave numbers of the Raman beams made to coincide with the lattice. This technique has previously been used in spectroscopic studies of atoms in optical lattices \cite{Stoferle2004,Kollath2006,Winkler2006,Massel2009,Tokuno2011,Poletti2011, Mark2011}. When analysing this version by considering time-dependent Wannier functions, we note that while the relationship between the lattice depth $V$ and the tunneling amplitude $J$ is nonlinear [i.e., a harmonic modulation of $V(t)$ will not produce a perfectly harmonic modulation of $J(t)$], the process can be made linear via a straightforward optimization can be used to determine a form for $V(t)$ that will produce a desired $J(t)$. Specifically, a modulation of $J$ at two different frequencies with amplitude $2 J_1$ can be produced via a specifically tailored anharmonic modulation of $V(t)$. The nature of modulating the lattice means that for a simple
modulation, we have $2|J_1| < |J_0|$, because otherwise we would have to switch the sign of $J$ in the modulation.

This procedure then gives rise to the same many-atom Hamiltonian described above, including the corresponding level of control over $\ueff$ and $\uefft$ by varying the modulation frequencies. We note the distinction between the lattice modulation methods we present here and the modifications to tunneling introduced by a time-varying potential gradient (AC driving field), which can be used to modify the magnitude of the tunneling constant \cite{Grossmann1991,Eckardt2005,Creffield2007,Lignier2007,Sias2008,Creffield2008,Gong2009,Hemmerich2010,Creffield2010} in a manner that depends on the amplitude and frequency of the drive. These schemes do have similarities, in so far as the effective interactions can be modified, and even tuned from repulsive to attractive by either technique \cite{Tsuji2011}. 

\begin{figure}[tb]
\includegraphics[width=7.5cm]{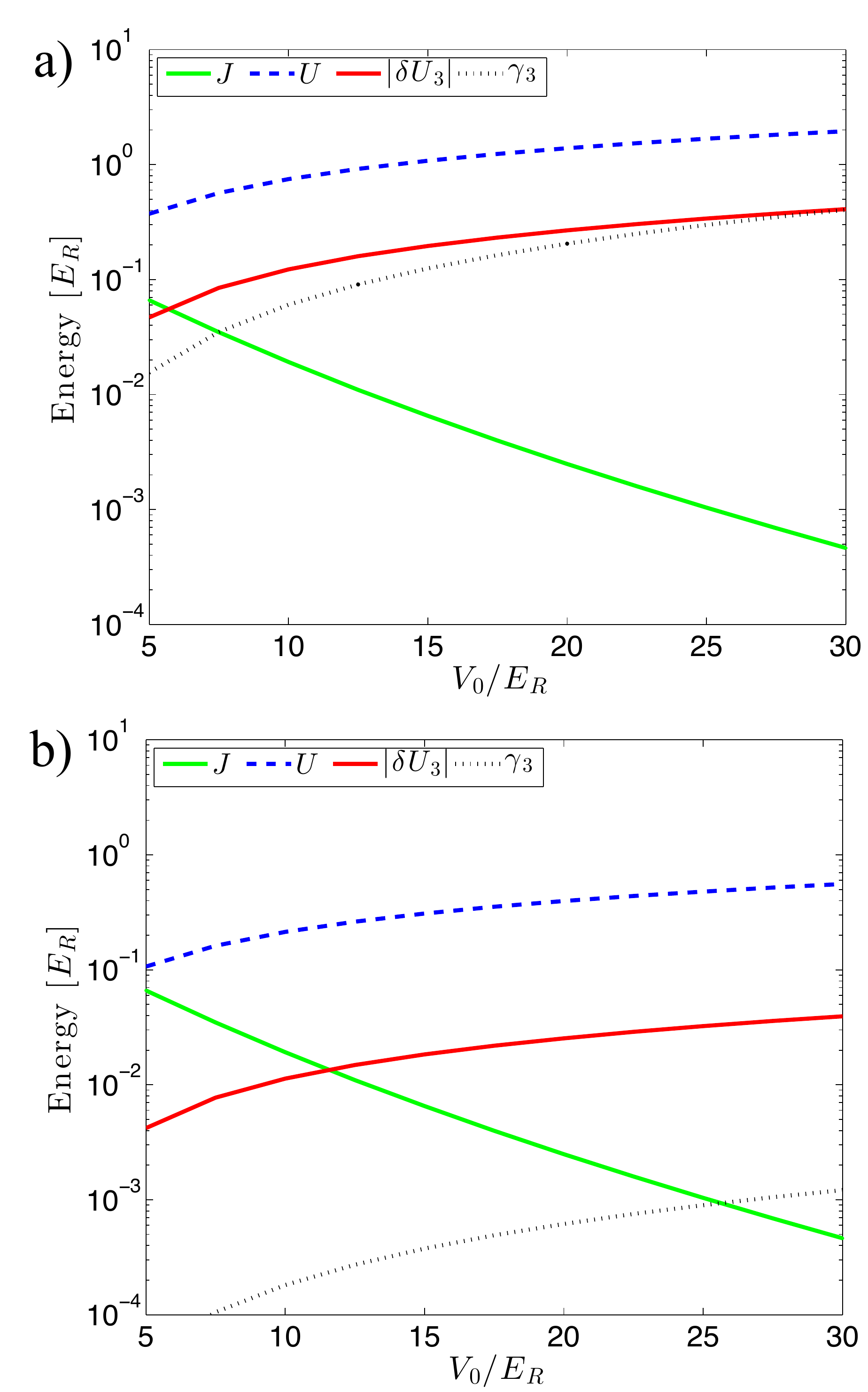}
\caption{Example parameter values for (a) $^{133}$Cs, with scattering length $a_s=350a_0$, where $a_0$ is the Bohr radius, and three-body parameter$L_3=0.5\times10^{-25}$cm$^6$s$^-1$; and (b) $^{87}$Rb, with scattering length $a_s=100a_0$, and $L_3=2.3\times10^{-28}$cm$^6$s$^{-1}$. Values are shown as a function of lattice depth $V_0$ in recoil energy $E_R$, and include the tunnelling $J$, unmodified on-site interaction $U$, three-body shift $\delta U_3$, and on-site three-body loss rate $\gamma_3$, estimated for atoms in the lowest band of an optical lattice as discussed in the text.} \label{fig:expparams}
\end{figure}

Naturally, the modulation will also cause variation of $U$ with time, $U=U_0+U_1(t)$. However, this variation occurs on a timescale also given by $U$, and is much faster than the dynamical timescales in the effective model. Under the same conditions as the rotating wave approximation that has already been made, this will therefore not alter the system dynamics. Outside of this limit other physical properties can be generated -- the extreme limit in which only $U$ is modulated was discussed recently in Ref.~\cite{Rapp2012}.
 
\subsection{Experimental parameters}

In Fig.~\ref{fig:expparams}, we show example parameters as a function of lattice depth, with typical numbers taken from $^{133}$Cs and $^{87}$Rb. In each case we note that at higher lattice depths, the system always reaches a limit where $|\delta U_3| \gg J$, so that the energy scale $U_3$ will dominate the dynamics. For comparison we have also plotted estimated three-body recombination rates for a triply occupied site, $\gamma_3$. For the Cs parameters, this scale is comparable to $\delta U_3$, as shown in Fig.~\ref{fig:expparams}a, so in order to prevent significant loss we would need to work in a limit where $\gamma_3, |\delta U_3| \gg J$. The resulting dynamics would be dominated by an effective loss-induced three-body constraint, as discussed in Ref.~\cite{Daley2009}. 

In this regime, the scheme described here further enhances the system lifetime via strong coherent three-body interactions $U_{\mathrm{eff}}^{(3)}$, and provides a new means to tune the effective two-body interaction $U_{\mathrm{eff}}^{(2)}$. In order to achieve a coherent three-body interaction it is necessary to also have small $\gamma_3$, which, e.g., could be achieved with $^{87}$Rb (see Fig.~\ref{fig:expparams}b). Because the scattering length for $^{87}$Rb is also typically smaller, reaching a regime where this scale is dominant requires large lattice depths where the timescales associated with $J$ can be relatively long. However, these regimes are very accessible for experiments with lighter species, e.g., $^{23}$Na or $^7$Li. These timescales could also be enhanced for Rb by using short-wavelength lattices.

\section{Mean-field Phase diagram}
\label{sec:mfpd}

\begin{figure}[tb]
\includegraphics[width=8cm]{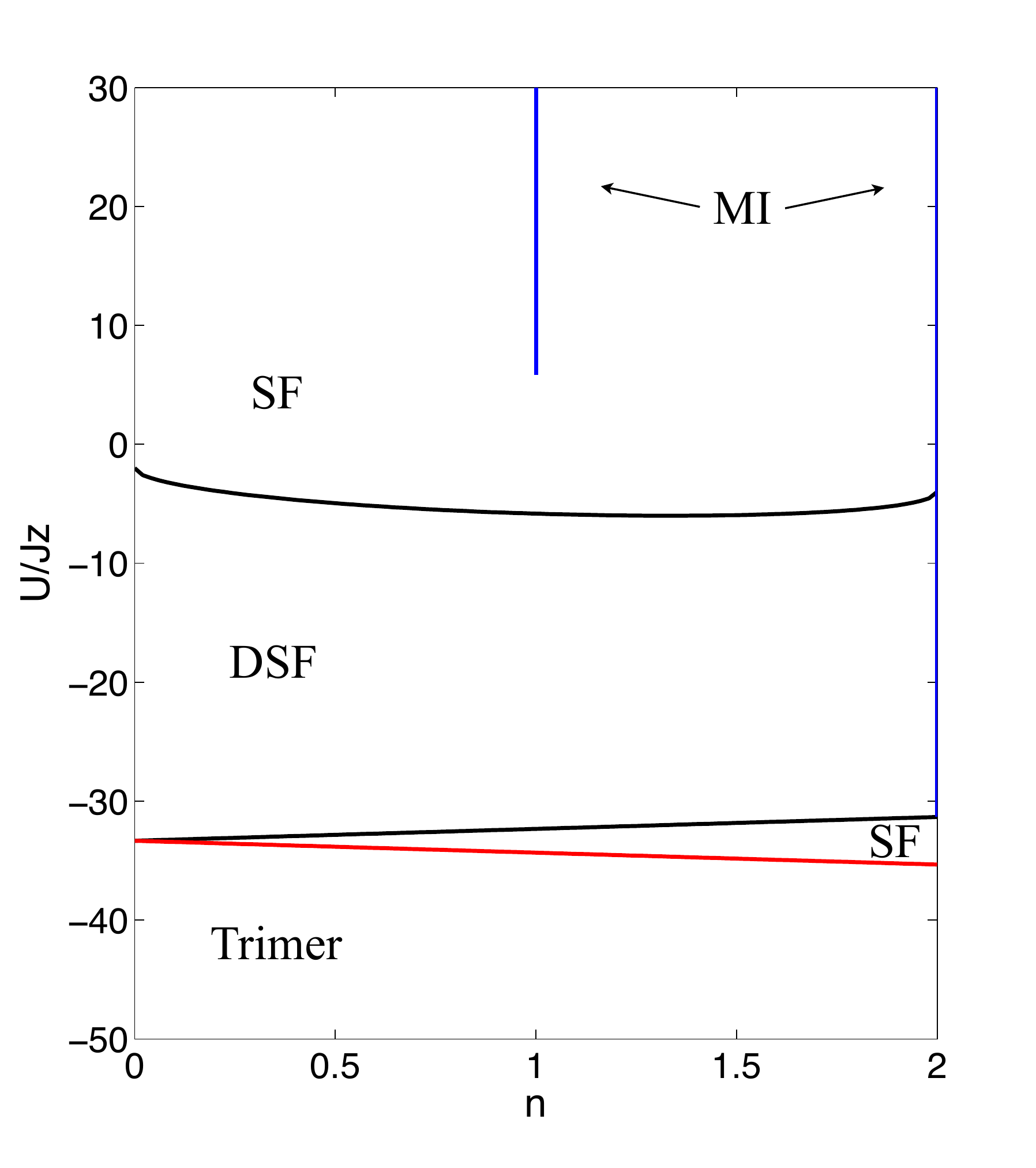}
\caption{Phase diagram from Gutzwiller mean-field theory for the Bose-Hubbard model with an additional effective three-body repulsion [Eq.~\eqref{effectiveham}], in a homogeneous system. We show the locations of the phase boundaries as a function of $\ueff/Jz$ and the mean filling factor $n$, and assume that the probability of occupation for sites with more than three particles is zero. Here we take $\uefft=50Jz$.} \label{fig:phasediag}
\end{figure}

To illustrate the physics accessible in this many-body model, we begin by deriving a mean-field phase diagram for the Hamiltonian Eq.~\eqref{effectiveham}. A Gutzwiller phase diagram was derived in Ref.~\cite{Daley2009} for the Bose-Hubbard model with a three-body hard-core constraint. Here we extend this phase diagram to the case of finite three-body
interactions (although with a four-body constraint), and also discuss the effects of the additional
single-particle tunneling term in $J_0'$.

Gutzwiller mean-field theory is formulated in terms of a product state ansatz, which takes the form $\ket\psi=\prod_{i}f_{n}|n_{i}\rangle$ for homogeneous systems. Here, $\ket{n_i}$ are states of well-defined on-site occupation on a single site, $i$, and $f_{n}$ are the coefficients of the different occupations. We find the state that minimizes the energy subject to normalization constraints, and a fixed mean particle number $n=\langle \hat n_i \rangle$. 

The phases that we obtain here are mostly similar to those obtained in the case of a three-body hard-core constraint \cite{Daley2009}. They are:
(i) The Superfluid (SF) phase [sometimes referred to in this model as an atomic superfluid (ASF)],
characterized by off-diagonal long-range order for single atoms {[}and single-atom superfluidity{]}, or algebraic decay of off-diagonal order
in 1D, which we refer to as quasi-long-range order; (ii) the Mott Insulator (MI) phase, where gaps open in the spectrum that are induced by interactions, which is characterized in the Gutzwiller calculations by fixed mean density, and zero superfluid order parameter $\langle b_i \rangle=0$; and (iii) the dimer superfluid (DSF), where single atoms no longer exhibit long-range order or superfluidity, but rather, atoms are paired, and we observe off-diagonal long range order {[}quasi-long range in 1D{]} and superfluidity for dimers.
In the Gutzwiller mean-field approach, the effective order parameter for the superfluid phase is $\langle b\rangle,$ which is zero in
the MI phase, and non-zero in the SF phase. Similarly, for dimer superfluidity, we require $\langle b^{2}\rangle$
to be non-zero, but $\langle b \rangle=0$. 

The mean-field phase diagram can be obtained analytically, and is plotted in Fig.~\ref{fig:phasediag} for $J_0=J_1\equiv J$, and $\uefft=50J$, for varying $n$ and $\ueff/J$. for $\ueff>0$ we observe the standard occurrence of the Mott Insulator phase at integer $n$, and otherwise the system is superfluid. The MI-SF transition for $n=1$ occurs at the standard value of $\ueff/(Jz)=3+2\sqrt{2} $, where $z$ is the coordination number for each lattice site. The three-body interaction ensures that the Mott Insulator phase for $n=2$ extends to attractive values of $\ueff$. For $\ueff<0$, we observe a DSF at most filling factors, with the transition for $n=1$ occurring at $\ueff/(Jz)=-3-2\sqrt{2}$. As $\ueff$ is further increased, we see the effects of the finite three-body interaction. We first observe a transition where the DSF is broken because it is favorable to create a mixture of doubly and triply occupied sites, and then another transition where it is not favorable to occupy any site with less than three particles, creating a phase of trimers.

We note that when $J_0' \neq 0$, there will be shifts of this phase diagram. Even for $\ueff=0$ there is a significant difference in the Gutzwiller coefficients for the two tunneling terms, and it is possible that $J_0' \neq 0$ may give rise to an additional type of ordering not captured within mean-field.

\section{Controlling dimer-dimer interactions for attractive bosons}
\label{sec:dimerint}

Along the $n=1$ line for large $|\ueff | \gg J_1$, with $\ueff<0$
we expect that there can be a competition between DSF order and charge-density wave (CDW)
order \cite{Schmidt2009,Diehl2010a,Diehl2010b,Diehl2010d}.  The latter is characterised by diagonal order, i.e., a crystal-like state with order in density-density correlation functions $\langle \hat n_i \hat n_j \rangle -\langle \hat n_i \rangle\langle \hat n_j \rangle$ (although in 1D with algebraically decaying correlations).  In this regime, essentially all particles are bound as dimers, and
we can write an effective model in second order perturbation theory
for the motion of dimers, which behave as hard-core bosons with operators
$d_{i}=b_{i}^{2}$ \cite{Schmidt2009,Diehl2010a,Diehl2010b,Diehl2010d}. This is based on restricting the system to a manifold
in which all particles are bound dimers, and computing coupling between states via virtual transitions to other manifolds. The corresponding Hamiltonian is 
\begin{equation}
H_{d}=\sum_{\langle i,j\rangle}\left(-J_{d}d_{i}^{\dagger}d_{j}+U_{d}d_{i}^{\dagger}d_{i}d_{j}^{\dagger}d_{j}+ 2\varepsilon_j d_j^\dagger d_j \right)\label{eq:dimerham}
\end{equation}
Here, the coefficients $J_{d}$ and $U_{d}$ can be computed in second order
perturbation theory, and depend on how the dimers are formed, and on effective three-body interactions. 
For notational convenience, let us write $J\equiv J_1$. We note that in second order perturbation theory the terms in $J_0$ play no role, because all processes appearing in this order involve breaking dimers.

\begin{figure}[tb]
\includegraphics[width=7.5cm]{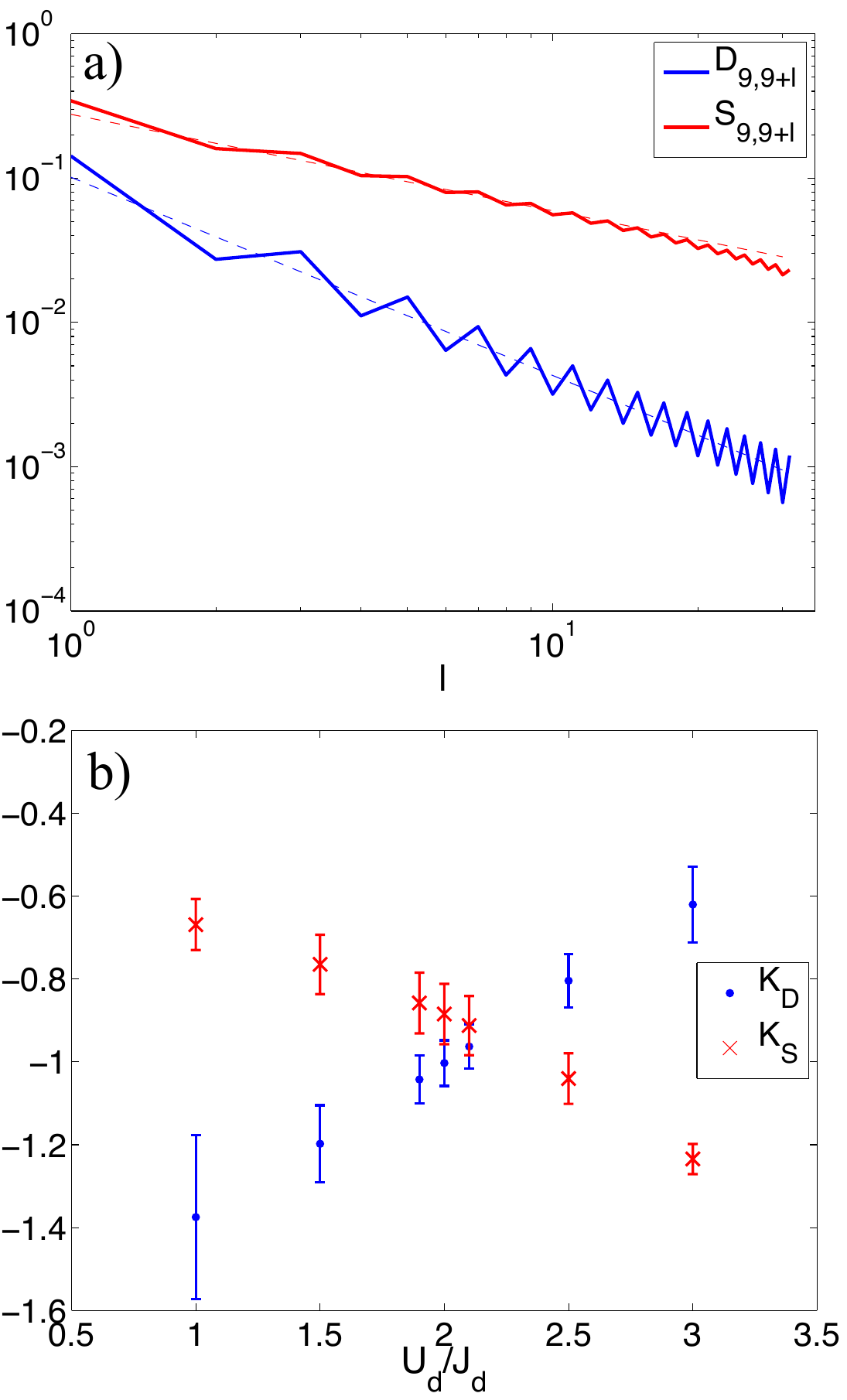}
\caption{Ground state correlations in the interacting dimer model (eq.~\eqref{eq:dimerham}), for $24$ dimers on $48$ lattice sites, with box boundary conditions and $\varepsilon_i =0$. (a) Example correlation functions $S_{i,j}=\langle d^\dagger_i d_j \rangle$ (which indicates DSF order), and $D_{i,j}=\langle  d^\dagger_i d_i  d^\dagger_j d_j  \rangle - \langle  d^\dagger_i d_i \rangle \langle  d^\dagger_j d_j  \rangle$ (which indicates CDW order) shown for $U_d=J_d$. The thin dashed lines show linear fits on the double-logarithmic scale. (b) Exponents for algebraic decay of the correlation functions $D_{i,j}$, $K_D$, and $S_{i,j}$, $K_S$. These are determined by fitting to correlation functions starting from site $9$ (to avoid boundary effects), and extending over 15 lattice sites. Error bars represent the range of $K$ values fitted by least squares methods to the correlation functions $S_{9,9+l}$ and $D_{9,9+l}$  points over $l=$7--15 sites. These calculations are converged with DMRG bond dimension $\chi=200$.} \label{dimercorr}
\end{figure}

As demonstrated in Fig.~\ref{dimercorr}, the competition between DSF and CDW order in a 1D system can be characterized by the algebraic decay of the single dimer density matrix, $S_{i,j}=\langle d^\dagger_i d_j \rangle$ (which indicates DSF order), and the density-density correlation function $D_{i,j}=\langle  d^\dagger_i d_i  d^\dagger_j d_j  \rangle - \langle  d^\dagger_i d_i \rangle \langle  d^\dagger_j d_j  \rangle$ (which indicates CDW order). At long distances, the physics will be dominated by the order for which the corresponding correlation function has the slower algebraic decay. In Fig.~\ref{dimercorr}a, we show the correlation functions for $J_{d}=U_{d}$ and a system at half filling of dimers, which are computed in imaginary time evolution with time-dependent density matrix renormalization group (t-DMRG) methods \cite{Vidal2003,Vidal2004,Daley2004,White2004,Verstraete2008}. In Fig.~\ref{dimercorr}b, we show extracted decay exponents $K_S$ and $K_D$ for the correlation functions, together with fitting errors. In good agreement with expressions obtained previously analytically \cite{Schmidt2009,Diehl2010a,Diehl2010b,Diehl2010d} we find that in a 48-site system  with half-filling of dimers, DSF order dominates when $U_d \gtrsim 2J_d$. We note that for open boundary conditions, it is important to calculate correlation functions in the center of the system, as the correlation functions will decay faster due to boundary effects. In a harmonic trap the filling will very across the system, and different phases can dominate in different regions, as is discussed in Ref.~\cite{Diehl2010b}.

In the case discussed in Refs.~\cite{Diehl2010a,Diehl2010b,Diehl2010d}, with a three-body hard-core constraint, 
\[
J_{d}=\frac{2J^{2}}{U_{\mathrm{eff}}^{(2)}},\, U_{d}=\frac{4J^{2}}{U_{\mathrm{eff}}^{(2)}},
\]
which leaves the system exactly on the boundary where the CDW and DSF orders
coexist.  With attractively bound dimers, and no three-body constraint (just
metastability) \cite{Schmidt2009},
\[
J_{d}=\frac{2J^{2}}{U_{\mathrm{eff}}^{(2)}},\, U_{d}=\frac{8J^2}{U_{\mathrm{eff}}^{(2)}},
\]
where interactions dominate, and only the CDW appears. Here, with a finite 3-body interaction $U_{3}$ that dominates $U$, we
find instead that 
\[
J_{d}=\frac{2J^{2}}{U_{\mathrm{eff}}^{(2)}},\, U_{d}=\frac{4J^{2}}{U_{\mathrm{eff}}^{(2)}}-\frac{4J^{2}}{\uefft}.
\]
The sign of the correction, and thus the determination of whether we observe CDW or DSF order is based on the sign of $\uefft$. For $\uefft>0$, DSF order is favored, and for $\uefft<0$, CDW order is favored.

The CDW phase could be most easily observed in experiments with a quantum gas microscope, directly measuring the distribution of doubly-occupied sites \cite{Bakr2011} and computing the related correlation functions. The DSF phase is characterised by dimers with long-range order. In an experiment, one would see (1) pairs forming (i.e., a large probability
of observing two atoms per site), and (2) as the DSF forms from a
SF phase, disappearance of the interference peak in the single-particle
momentum distribution. This is because the long range order for single
atoms goes away. To see long range order for dimers, the dimers could
be associated to molecules and then released from the lattice to perform
a time-of flight measurement of the momentum distribution.  In principle, this could be done with Feshbach molecules, at the risk
of some collisional loss from colliding molecules during the measurement
process. Another alternative would be to detect the dimer-dimer correlation functions using noise-correlation measurements \cite{Folling2005,Altman2004}. We discuss the adiabatic preparation of states in these regimes below.

\section{Metastable phases and adiabatic state preparation}

\label{sec:prepare}

\begin{figure}[tb]
\includegraphics[width=8.5cm]{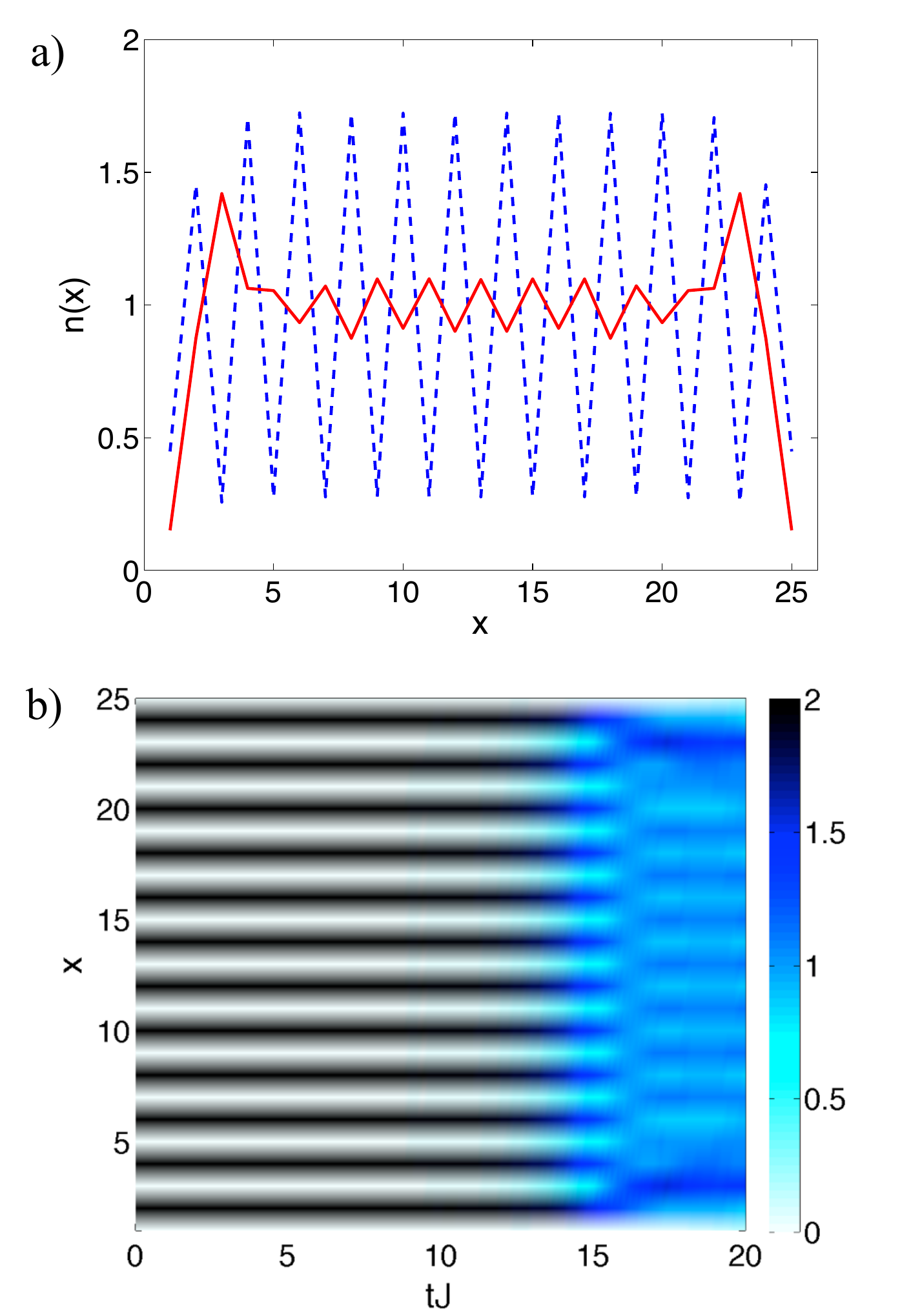}
\caption{Adiabatic preparation of dimer states from a Mott insulator in a superlattice. Here we show the results of time-dependent DMRG numerical simulations based on eq.~\eqref{effectiveham}, showing a time-dependent ramp in a superlattice of period 2, from a two-particle Mott insulator in the lowest wells to regimes with different signs of $U_{\rm eff}^{(3)}$, which change the strength of the effective off-site interaction between dimers. The superlattice depth is ramped from $V_{SL}=20J$ to $V_{SL}=0$ in an exponential ramp with a time constant $\tau_{\rm ramp}=4J^{-1}$. We choose $U_{\rm eff}^{(2)}=-8J$ and compute the dynamics for 24 particles on 25 lattice sites with box boundary conditions. (a) Mean on-site occupation $\langle \hat n_x \rangle$ at the end of the ramp for occupation $U_{\rm eff}^{(3)}>0$ (red solid line) and $U_{\rm eff}^{(3)}<0$ (blue dashed line). We observe enhanced oscillations characteristic of the stronger repulsive interaction generated when $U_{\rm eff}^{(3)}<0$ (see the text for details). (b) The on-site occupation $\langle \hat n_i \rangle$ plotted as a function of time during the ramp for $U_{\rm eff}^{(3)}>0$. } \label{fig:figure1d}
\end{figure}

\begin{figure}[h]
\includegraphics[width=7cm]{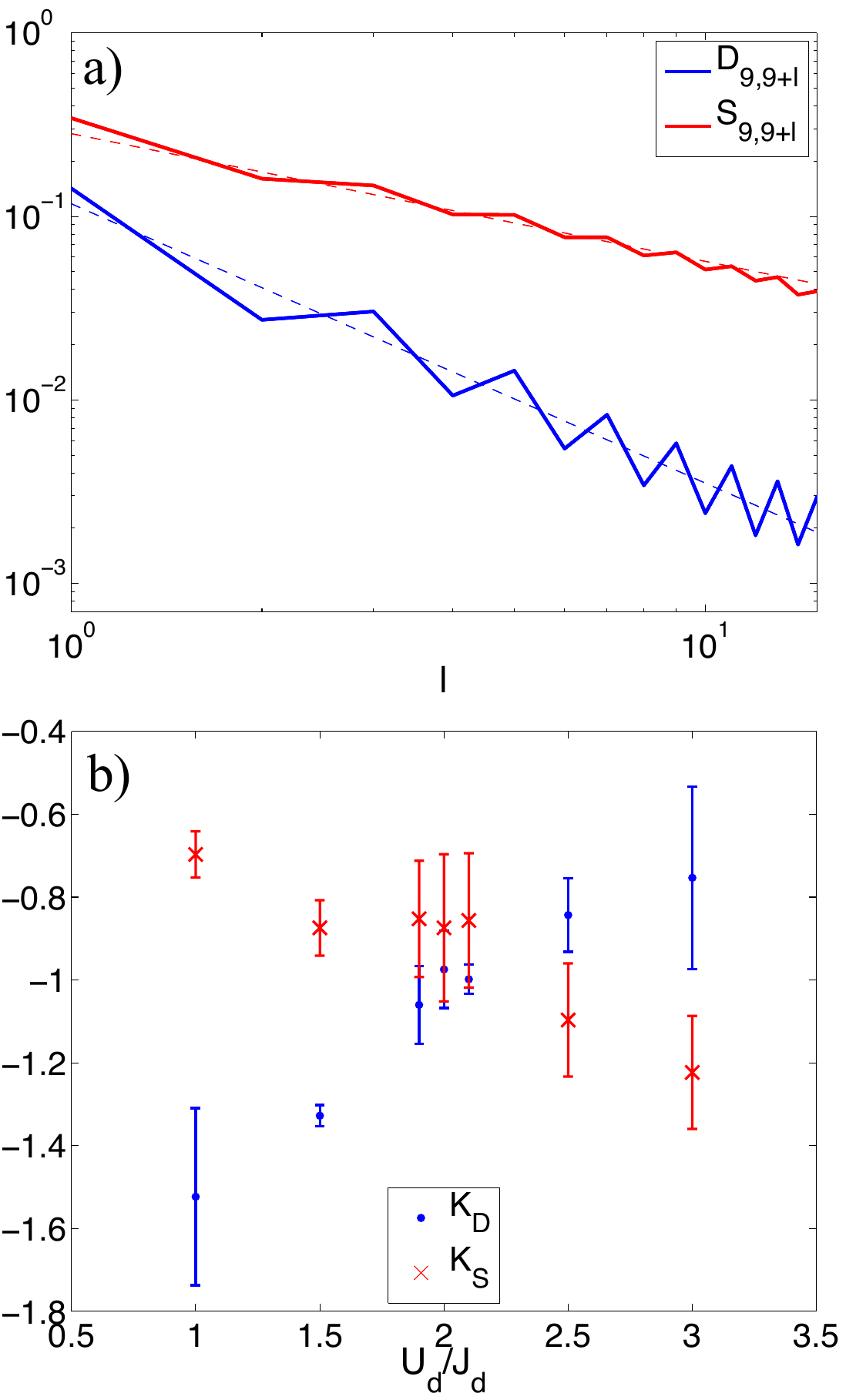}
\caption{Adiabatic preparation of states with CDW and DSF order from a Mott insulator in a superlattice. Here we show the results of t-DMRG numerical simulations based on eq.~\eqref{eq:dimerham}, showing a time-dependent ramp in a superlattice of period 2, from a Mott insulator with one dimer in each of the lowest wells, for varying $U_d$. The superlattice depth is ramped from $V_{SL}=20J_d/2$ to $V_{SL}=0$ in an exponential ramp with a time constant $\tau_{\rm ramp}=10J_d^{-1}$, with $24$ dimers on $48$ lattice sites and box boundary conditions. The correlation functions shown are from the end of a ramp after a time $T=80J_d^{-1}$, with exponential ramp divided into timesteps of time $\delta t =0.1 J^{-1}$ with constant superlattice depth in each step. (a) Example correlation functions $S_{i,j}=\langle d^\dagger_i d_j \rangle$ (which indicates DSF order), and $D_{i,j}=\langle  d^\dagger_i d_i  d^\dagger_j d_j  \rangle - \langle  d^\dagger_i d_i \rangle \langle  d^\dagger_j d_j  \rangle$ (which indicates CDW order) shown for $U_d=J_d$. The thin dashed lines show linear fits on the double-logarithmic scale. (b) Exponents for algebraic decay of the correlation functions $D_{i,j}$, $K_D$, and $S_{i,j}$, $K_S$. These are determined by fitting to correlation functions starting from site $9$ (to avoid boundary effects), and extending over 11 lattice sites. Error bars and numerical parameters are the same as those for Fig.~\ref{dimercorr}.
} \label{dimercorrev}
\end{figure}

In a typical experiment, $U_{\mathrm{eff}}^{(3)}$ will be negative irrespective of the value of $U$, because it arises in second order perturbation theory from coupling of atoms in the lowest band to higher bands. Although RF couplings to 3-body bound states can give rise to positive shifts \cite{Safavi-Naini2012}, and hence $U_{\mathrm{eff}}^{(3)}>0$, it is therefore important to ask how interesting states can be prepared without the system collapsing due to the attractive interactions. For this, we need to consider the case where we form metastable many-body states, by beginning in states that contain essentially no triply occupied sites and allowing the strong attractive three-body interaction to prevent triple occupation. This works in an optical lattice because there on typical experimental timescales there are no mechanisms available to dissipatively remove energy from the system, and triply occupied sites therefore cannot be formed due to energy conservation. Analogous metastable states have already been realised in experiments with atoms in optical lattices, including repulsively bound atom pairs \cite{Winkler2006,Strohmaier2010}, Ising models in a tilted lattice \cite{Simon2011,Sachdev2002,Pielawa2011,Meinert:2013aa}, Metastable Mott Insulator states \cite{Mark2012}, and states with negative temperature \cite{Braun:2013aa}. 

These states can be prepared, e.g., via adiabatic state preparation, starting in a gapped phase that can be prepared with low entropy (e.g., a band insulator in a superlattice), which has no triply occupied sites. We then vary the Hamiltonian parameters in time to transfer the state to a more complicated many-body phase that corresponds to an eigenstate of another Hamiltonian \cite{Rabl2003, Trebst2006,Kantian2010,Sorensen2010}.

An example in the context of the present scheme would be a process in which the interactions are gradually altered after beginning in a Mott Insulator state at unit filling. In this scenario, the detuning of the couplings, $\omega_{m1}$ and $\omega_{m2}$ would be changed gradually to introduce the coupling. This would be done by switching on the photon-assisted tunnelling with initial coupling frequencies that are very far off the resonance (at $U$), and then changing the
coupling frequency so that $U_{\mathrm{eff}}^{(2)}$ goes from being
initially large and positive (the MI regime, $U_{\mathrm{eff}}^{(2)} \gg J_0$ ) to being small, and potentially
negative (with $|U_{\mathrm{eff}}^{(2)}| \sim J_0, J_1$ . Throughout, $\left|U_{\mathrm{eff}}^{(3)}\right|$ will
be larger than the other system parameters, and hence three-body occupation will remain small. But in the regime $U_{\mathrm{eff}}^{(2)} \sim J_0, J_1$, it is then possible to realize the superfluid and dimer superfluid phases depicted in Fig.~\ref{fig:phasediag}.

An alternative process is to begin in a Mott insulator state in the presence of a superlattice potential, where each of the lowest energy wells is occupied with an atom, and all other sites are unoccupied. In this state, the filling factor is controlled by the periodicity of the superlattice, and the detuning of the photon-assisted coupling does not need to be ramped during the preparation process. The superlattice can then be time-dependently modified to allow the particles to delocalize, with the final filling factor chosen via the superlattice period. An example of this process for formation of dimer states is discussed below.

The use of adiabatic state preparation has added importance here, in that it can help to produce states of interest with low entropies. While the temperatures required to produce superfluid and Mott insulator states with repulsive two-body interactions $\ueff$ will be similar to those in previous experiments, realizing a superfluid of dimers will be more difficult. The energy scales associated with dimer tunneling, $~J_1^{2}/U_{\mathrm{eff}}^{(2)}$ are much smaller, leading to lower required temperatures, as was confirmed in 2D recently from numerical simulations
of a Bose-Hubbard model with three-body constraint \cite{Bonnes2011}. However, such states could be formed via adiabatic state preparation in initial states of very low entropy, as was demonstrated for the 1D Ising model in a tilted lattice \cite{Simon2011,Meinert:2013aa}.

\subsection{Time-dependent adiabatic state preparation in 1D}

We illustrate adiabatic state preparation by studying the formation dynamics of the dimer superfluid states discussed in section \ref{sec:dimerint}. Here we consider a scheme similar to that proposed in Ref.~\cite{Daley2009}, where we begin with a Mott insulator state in the presence of a superlattice potential, here of period 2. We begin in a lattice with $\varepsilon_i=V_{SL}$ for $i$ odd, and $\varepsilon_i=0$ for $i$ even. This superlattice is loaded with a state in the deep Mott insulator limit, $U\gg J$, where two atoms occupy all of the lower energy wells. Then the coupling terms are switched on to produce the effective interactions $\ueff$ and $\uefft$, and the superlattice is slowly removed in time, $\varepsilon \rightarrow 0$.

For 1D systems we can compute time-dependent dynamics of this ramp exactly for experimentally relevant timescales, parameters and system sizes. This is possible using t-DMRG methods \cite{Vidal2003,Vidal2004,Daley2004,White2004,Verstraete2008}, which provides a means to exactly propagate 1D many-body states exactly, provided that the dynamics are not too far from equilibrium. Corresponding results are shown in Fig. \ref{fig:figure1d}, Where we consider a large value of $|\uefft|=50J$ and $\ueff=-8J$, for $\uefft>0$ and $\uefft<0$. In Fig.~\ref{fig:figure1d}a, we show the on-site density at the end of the ramp, where we see the strong increase in dimer-dimer interactions in the case $U_{\rm eff}^{(3)}<0$ compared with $U_{\rm eff}^{(3)}>0$, as is expected from the perturbation theory discussion in section \ref{sec:dimerint}. In Fig.~\ref{fig:figure1d}b we show the time-dependence of this mean occupation during the adiabatic ramp.

In Fig.~\ref{dimercorrev}, we investigate this further in a larger system with a half filling of dimers, computing the dynamics described by eq.~\eqref{eq:dimerham} with $24$ dimers on $48$ lattice sites during an adiabatic ramp beginning in a superlattice with $V_{SL}=20J_{d}/2$. We find that the correlation functions $S_{i,j}$ and $D_{i,j}$ exhibit similar behaviour to that shown in Fig.\ref{dimercorr}, however for shorter ramp times that are experimentally accessible, the correlation functions exhibit algebraic decay only over a finite distance, which increases with increasing ramp time (as longer ramp times allow correlations to be established over longer distances). Here we choose a ramp time of $T=80J_d^{-1}$. At the end of the ramp, we fit exponents of the algebraic decay to the correlation functions over $11$ sites, including up to $15$-site correlation functions in our error estimates. The error bars are larger than for the equivalent ground state calculations because of the shorter distances used to compute the correlation functions, but the decay exponents obtained here agree with the ground state values within these error estimates. This confirms that even for finite ramp times where excitations are produced during the preparation process, we still observe the same qualitative physics in terms of CDW and DSF order that we expect in the ground state, as exhibited by measurable correlation functions.

\subsection{Selecting charge density wave and superfluid phases}

In the dimer model defined above, we noted that the coefficients are given by
\[
J_{d}=\frac{2J^{2}}{U_{\mathrm{eff}}^{(2)}},\, U_{d}=\frac{4J^{2}}{U_{\mathrm{eff}}^{(2)}}-\frac{4J^{2}}{\uefft}.
\]
For repulsive interactions, and assuming that $\uefft<0$, this would imply that $U_d/J_d<2$, meaning that a superfluid of dimers will always be favored over a CDW in the ground state of the Hamiltonian. However, even with the condition $\uefft<0$, we can use adiabatic state preparation to realize a CDW. 

In the example given above, the dimers were started in the lower wells of the superlattice potential. However, we could also begin with the dimers in the higher-energy wells. This state is connected adiabatically to the \emph{highest-energy} state of the Hamiltonian $H_d$, or equivalently the ground state of the hamiltonian $-H_d$. Now consider the case where we tune the effective two-body interaction so that $U_{\mathrm{eff}}^{(2)}<0$. In this case, we see that 
\[
-J_{d}=\frac{2J^{2}}{\left|U_{\mathrm{eff}}^{(2)}\right|},\, -U_{d}=\frac{4J^{2}}{\left|U_{\mathrm{eff}}^{(2)}\right|}+\frac{4J^{2}}{\uefft}.
\]
So that the ground state of the Hamiltonian $-H_d$ will be dominated by CDW order.

Thus, by choosing the combination of the sign of $U_{\mathrm{eff}}^{(2)}$ and whether the dimers populate the high-energy sites or low-energy sites of the superlattice, we can select between a dimer superfluid and a CDW phase in the adiabatic preparation.

\section{Summary and Outlook}
\label{sec:summary}

In summary, we have introduced a method to control two-body interactions in an optical lattice in such a way that a dominant effective three-body interaction can be produced. This requires the engineering of a time-dependent tunneling term which could be created using laser-assisted tunneling or though lattice modulations. To illustrate the many-body properties that can be observed in this way, we discussed the mean-field phase diagram and state preparation in the presence of finite effective three-body interactions, which is similar in form to the case of hard-core three-body constraints in the region where three-body interactions are dominant. We also discussed adiabatic preparation of metastable phases in this system, including phases with CDW and superfluid order with tightly bound particle pairs. 

In contrast to previous proposals for observing such physics with a three-body constraint generated via three-body losses, the scheme here allows for longer adiabatic ramp times because the longest allowed timescales are not limited by real three-body losses. We also provide a separate mechanism to control two-body interactions in these systems, which may also be useful in its own right for species with large scattering lengths that are not easily controllable via Feshbach resonances (e.g., $^{86}$Sr).

These ideas could be combined with other recent proposals to enhance three-body interactions by coupling to three-body bound states \cite{Safavi-Naini2012} to make these interactions effectively stronger. They could also be extended to Fermions, where the case of three-component Fermions would be particularly interesting and relevant for ongoing experiments \cite{Cherng2007,Ottenstein2008,Rapp2007,Rapp2008}. In addition to the stabilization of dimer superfluid states predicted in Ref.~\cite{Kantian2009}, a finite-three body interaction is expected to result in interesting spin-ordering in these gases. 

\begin{acknowledgments}

We would like to thank Sebastian Diehl, Stephan Langer, Phillip Johnson, Manfred Mark, Hanns-Christoph N\"agerl, Hannes Pichler, Johannes Schachenmayer, Javier Von Stecker, and Eite Tiesinga for helpful and stimulating discussions. Work in Pittsburgh is supported by NSF Grant PHY-1148957, and computational resources were provided by the Center for Simulation and Modeling at the University of Pittsburgh. JS acknowledges support from the AFOSR and DARPA.
\end{acknowledgments}

   \def\eprint#1{arXiv:#1}

\bibliographystyle{apsrev}
\bibliography{daley_list}

\end{document}